\documentstyle[A4,12pt]{article}

\begin{document}
\newcommand{\un}{\underline}
\newcommand{\x}{\times}
\newcommand{\dd}{\ddagger}
\newcommand{\Ep}{{\cal E}}
\thispagestyle{empty}
\vspace*{2cm}
\begin{center}
{\large {\bf RACAH SUM RULE AND}}\\
\vspace*{5mm}
{\large {\bf BIEDENHARN-ELLIOTT IDENTITY}}\\
\vspace*{5mm}
{\large {\bf FOR THE SUPER-ROTATION $6-j$ SYMBOLS.}}
\end{center}
\vspace*{20mm}
\begin{center}
Pierre Minnaert and Stoyan Toshev*
\end{center}
\begin{center}Laboratoire de Physique Th\'{e}orique**, Universit\'{e} de
Bordeaux I, France***\end{center}
\vspace*{2cm}
\begin{center}{\bf Abstract}\end{center}

It is shown that the well known Racah sum rule and Biedenharn-Elliott identity
satisfied by the recoupling coefficients or by the $6-j$ symbols of the usual
rotation $SO(3)$ algebra can be extended to the corresponding features of the
super-rotation $osp(1|2)$ superalgebra. The structure of the sum rules is
completely similar in both cases, the only difference concerns the signs
which are more involved in the super-rotation case.

\vspace*{1cm}
\noindent
PACS.02.20 - Group Theory. \\
PACS.11.30P - Supersymmetry. \\

\vfill
\noindent
* On leave from Institute for Nuclear Research and Nuclear Energy,
Bulgarian Academy of Sciences, Sofia, Bulgaria.\\
{}~** Unit\'{e} associ\'{e}e au CNRS URA 764 \\
{}~~*** Postal address: Rue du Solarium, 33175 Gradignan CEDEX	\\
\\
{}~~LPTB-94-01 \hfill {January 1994}

\newpage
\section{Introduction}
\hspace*{10mm}The super-rotation algebra
also called graded $su(2)$ has been
investigated first by Pais and Rittenberg [1]. They have built explicitly
the finite dimensional representations of this algebra and have shown that
the irreducible representations are characterized by a superspin $j$,
which takes integer or half integer values, and by the parity $\lambda$
which takes the values $0$ or $1$. Later on,
Scheunert, Nahm and Rittenberg [2]
have introduced the concept of grade star representation and have computed
the Clebsch-Gordan coefficients that reduce the product of two super-rotation
representations. It turns out that these super-rotation Clebsch-Gordan
coefficients are products of the usual rotation Clebsch-Gordan coefficients,
which are independent of the parities, and a scalar factor that depends both
on the superspins and the parities of the representations. Berezin and
Tolstoy [3] have shown that these scalar factors form a pseudo-orthogonal
matrix and have derived the explicit matrix elements of the
$j$ representation in standard basis.

Recently, Minnaert and Mozrzymas [4] have shown that in order to reduce the
tensor product of two arbitrary representations one must introduce
the most general Hermitian form in the representation space.
The Racah-Wigner calculus for
the super-rotation algebra was then developed with a definition of the
super-rotation $S3-(j,\lambda)$ and $S6-(j,\lambda)$ symbols
and an analysis of their properties.
In subsequent papers [5,6], Daumens, Minnaert, Mozrzymas and Toshev  have
defined $\lambda$-parity independent super $S3-j$ and
$S6-j$ symbols and studied their symmetries.

It is well known from the usual rotation case [7] that
the study of the $6-j$ symbols is interesting for several reasons:
{\em (i)} since the $6-j$ symbols are
independent of the projection quantum numbers
they are basis independent objects ;
{\em (ii)} the $3-j$ symbols ( or, equivalently, the Clebsch-Gordan
coefficients )
can be obtained as a limit of the $6-j$ symbols; {\em (iii)} the  $6-j$
symbols are proportional to the recoupling coefficients for
the addition of three angular momenta and constitute a ``basis" for all
recoupling coefficients in the addition of an arbitrary number of
angular momenta.
In some sense, it is the $6-j$ symbols that are the basic objects
of the angular momentum theory, since all other entities can be derived
from them. In turn, the rotation $6-j$ symbols are determined
up to a phase convention by the orthogonality relations,
the Racah sum rule, the Biedenharn-Elliott identity and their symmetry
properties.

Therefore, carrying on the development of the Racah-Wigner
calculus for the super-rotation algebra, we derive in the present paper
the fundamental relations satisfied by the super-rotation $S6-j$
symbols, namely {\em the Racah sum rule} and
{\em the Biedenharn-Elliott identity}.
It will be shown that, similarly to the (pseudo-) orthogonality
relations already discussed in Refs. [4,5],
the Racah sum rule and the Biedenharn-Elliott
identity for the $S6-j$ symbols possess the same structure as the
corresponding relations for the rotation algebra apart from the
signs which are more involved in the super-rotation case, see
{\em e.g.} eqs. (3.6)-(3.7) and (4.3)-(4.4) below.

\section{The Super-Rotation Algebra}
\hspace*{10mm}The {\em super-rotation algebra}
defined by Pais and Rittenberg [1] has
the following (anti)\-commutation relations
  \begin{eqnarray}
 [J_i\:,\:J_j] &=& \imath \,\epsilon_{ijk} \: J_k \,, \\
\left [J_i\:,\:J_\alpha \right] &=& {\frac{1}{2}} J_\beta \:
(\sigma_i)^\beta_\alpha \,, \\
\left \{J_\alpha\:,\:J_\beta \right \} &=& {\frac{1}{2}} J_i \:
(\Gamma\sigma_i)_{\alpha\beta} \,,
  \end{eqnarray}
where $i,j,k=1,2,3$ , $\alpha , \beta=\pm {\frac{1}{2}}$ , $\sigma_i$ are the
Pauli matrices and $\Gamma=\imath \,\sigma_2$ .
This su\-per\-al\-ge\-bra with complex structure constants is related to the
orthosymplectic superalgebra $osp(1\mid 2)$ [2-5].

An irreducible representation of this algebra [1-4] is characterized
by the {\em superspin} $j \,(j=0, 1/2, 1,...)$
and the {\em parity} $\lambda \,(\lambda=0,1)$
which is the Grassmann degree of the highest weight vector in the graded
representation space. The superspin $j$
multiplet contains two rotation multiplets with spins $\ell=j$
and $\ell=j-1/2$.
Its dimension is $1$ if $j=0$ and $4j+1$ if $j \neq 0$.
The standard basis in the space $V(j,\lambda)$ is denoted by $\mid
j,\lambda;\ell,m>$.

In order to reduce the tensor
product of two arbitrary representations, it is necessary to consider the most
general pseudo-Hermitian form in $V(j,\lambda)$ [4]
 \begin{equation}
\Phi^{j,\lambda}_{\varphi\psi}(j,\lambda;\ell,m\mid
j,\lambda;\ell',m')=(-1)^{k\varphi+\psi}\,\delta_{\ell\ell'}\,\delta_{mm'}\,,
 \end{equation}
where $\varphi$ and $\psi$ are two binary variables that can take
the values $0$ or $1$ (mod $2$) such that
$\lambda+\varphi+\epsilon=1 \pmod{2}$.
The representation space $V(j,\lambda)$ in which acts an irreducible
representation of class $\epsilon$ and equipped with the bilinear Hermitian
form  $\Phi^j_{\varphi\psi}$ will be denoted by ${\cal H}^{j,\lambda}_
{\varphi\psi}$.
If the bilinear Hermitian form $\Phi$ in
the tensor product space
${\cal H}^{j_1,\lambda_1}_{\varphi_1\psi_1}\otimes{\cal H}^{j_2,\lambda_2}_
{\varphi_2\psi_2}$
is defined by
 \begin{equation}
\Phi(x_1 \otimes x_2\,,y_1 \otimes y_2)=(-1)^{\alpha(x_2) \alpha(y_1)}
\:\Phi^{j_1,\lambda_1}_{\varphi_1\psi_1}(x_1,y_1)
\:\Phi^{j_2,\lambda_2}_{\varphi_2\psi_2}(x_2,y_2)\,,
 \end{equation}
then [2] the tensor product of {\em grade star}
representations of class $\epsilon$ is also a grade star representation  of
same class and it is
fully reducible ( $\alpha(x_2)$ and $\alpha(y_1)$ are the Grassmann
degrees of the vectors $x_2$ and $y_1$, respectively ).
The reduction formula for the tensor product of the spaces reads
 \begin{equation}
{\cal H}^{j_1,\lambda_1}_{\varphi_1\psi_1}\otimes
{\cal H}^{j_2,\lambda_2}_{\varphi_2\psi_2}
=\bigoplus_{j_{12}}{\cal H}^{j_{12},\lambda_{12}}_{\varphi_{12}\psi_{12}}\,.
 \end{equation}
The main difference with the usual rotation case resides in the fact that the
possible values of $j_{12}$ in the direct sum are
\begin{equation}
j_{12}=\mid j_1-j_2 \mid , \mid j_1-j_2 \mid +\frac{1}{2},...,
j_1+j_2- \frac{1}{2}, j_1+j_2.
 \end{equation}
i.e., they vary with a step of $1/2$ and not $1$. As a consequence, the sum
of the three superspins, $j_1, j_2, j_{12}$ is not
necessarily an integer, it can be half-integer as well.

The basis
vectors in the space ${\cal H}^{j_{12},\lambda_{12}}_{\varphi_{12}\psi_{12}}$
are denoted by
$\mid (j_1 \lambda_1;j_2 \lambda_2) j_{12} \lambda_{12};\ell_{12}
m_{12}\rangle$.
They are related to the tensor product basis by the {\em super-rotation
Clebsch-Gordan coefficients} (SRCG)
 \begin{eqnarray}
& &\mid (j_1 \lambda_1;j_2 \lambda_2)
j_{12} \lambda_{12};\ell_{12} m_{12}\rangle \nonumber\\
& &=\sum_{\ell_1m_1\ell_2m_2}\mid j_1\lambda_1;\ell_1 m_1\rangle\otimes\mid
j_2 \lambda_2;\ell_2 m_2\rangle
(j_1 \lambda_1;\ell_1 m_1;j_2 \lambda_2;\ell_2 m_2\mid
j_{12} \lambda_{12};\ell_{12} m_{12})\,.\;\;\;\;\;
 \end{eqnarray}
The SRCG have been computed in [2,3] where it was shown that they can be
factorized into two factors
 \begin{eqnarray}
& &(j_1 \lambda_1;\ell_1 m_1;j_2 \lambda_2;\ell_2 m_2\mid
j_{12} \lambda_{12};\ell_{12} m_{12}) \nonumber\\
& &=\left( \begin{array}{cc}
j_1\lambda_1&j_2\lambda_2\\
\ell_1	    &\ell_2	 \\
\end{array}\right.
\left\|\begin{array}{c}
j_{12}\lambda_{12} \\
\ell_{12} \\
\end{array}\right)
(\ell_1m_1\ell_2m_2\mid\ell_{12}m_{12})\,
 \end{eqnarray}
The second factor is the usual rotation Clebsch-Gordan
coefficient that is independent of the parities whilst the first factor,
called the {\em scalar factor} because it is independent
of the magnetic quantum numbers,
depends on the parities of the representations.
The main point, stressed in [4],
is that the scalar factors satisfy pseudo-orthogonality relations which are
easily understandable in terms of pseudo-Hermitian forms that
satisfy relations (2.5).
Of course, since the usual rotation Clebsch-Gordan coefficients are
orthonormalized, the SRCG satisfy pseudo-orthogonality relations
similar to those of the scalar factors [4,5].

{}From eq.(2.8), using (2.5) and the symmetry properties of the super-rotation
Clebsch-Gordan coefficients derived in [4]:
 \begin{eqnarray}
&&(j_1 \lambda_1;\ell_1 m_1;j_2 \lambda_2;\ell_2 m_2\mid
j_{12} \lambda_{12};\ell_{12} m_{12}) \nonumber\\
&&=(-1)^{\ell_1+\ell_2-\ell_{12}+
(\lambda_1+I_{12})(k_2+I_{12})+(\lambda_2+I_{12})(k_1+I_{12})} \nonumber\\
&&\qquad\times(j_2 \lambda_2;\ell_2 m_2;j_1 \lambda_1;\ell_1 m_1\mid
j_{12} \lambda_{12};\ell_{12} m_{12})
 \end{eqnarray}
we obtain the following useful
formula concerning the symmetry properties of the basis vectors
in ${\cal H}^{j_{12},\lambda_{12}}_{\varphi_{12}\psi_{12}}$:
 \begin{eqnarray}
& &\mid (j_1 \lambda_1;j_2 \lambda_2)
j_{12} \lambda_{12};\ell_{12} m_{12}\rangle \nonumber\\
& &=(-1)^{(\lambda_1+I_{12})(\lambda_2+I_{12})+[j_1+j_2-j_{12}]}
\mid (j_2 \lambda_2;j_1 \lambda_1)
j_{12} \lambda_{12};\ell_{12} m_{12}\rangle.
 \end{eqnarray}
where $[j_1+j_2-j_{12}]$ is the integer part of $j_1+j_2-j_{12}$.

Let us consider the reduction of the tensor product
of three representations $(j_i,\lambda_i), i=1, 2, 3 $. This reduction can be
done in two different ways. Either, one couples first the representations
$(j_1,\lambda_1)$ and $(j_2,\lambda_2)$ and then the result
$(j_{12},\lambda_{12})$
is coupled to the representation $(j_3,\lambda_3)$ in order to give as a final
result the representation $(j,\lambda)$. Or, one couples $(j_1,\lambda_1)$
with the result $(j_{23},\lambda_{23})$ of the coupling of the
representations $(j_2,\lambda_2)$ and $(j_3,\lambda_3)$
in order to yield $(j,\lambda)$.
The corresponding basis vectors denoted by
$\mid ((j_1 \lambda_1;j_2 \lambda_2)j_{12} \lambda_{12};
j_3 \lambda_3) j \lambda;\ell m\rangle$ and
$\mid (j_1 \lambda_1;(j_2 \lambda_2;j_3 \lambda_3)
j_{23} \lambda_{23}) j \lambda;\ell m\rangle$ are
related by the so called {\em super-rotation recoupling
coefficients} [4]
 \begin{eqnarray}
& &\mid (j_1 \lambda_1;(j_2 \lambda_2;j_3 \lambda_3)
j_{23} \lambda_{23}) j \lambda;\ell m\rangle \nonumber\\
& &=\sum_{j_{12}} [(j_1 \lambda_1;(j_2 \lambda_2;j_3 \lambda_3)
j_{23} \lambda_{23}) j \lambda
\!\mid((j_1 \lambda_1;j_2 \lambda_2)j_{12} \lambda_{12};
j_3 \lambda_3) j \lambda] \nonumber\\
& &\qquad\times\mid ((j_1 \lambda_1;j_2 \lambda_2)
j_{12} \lambda_{12};
j_3 \lambda_3) j \lambda;\ell m\rangle.
 \end{eqnarray}
which are equal, up to a sign, to the {\em parity dependent
super-rotation} $S6-(j,\lambda)$ {\em symbols}

 \begin{eqnarray}
& &[(j_1 \lambda_1;(j_2 \lambda_2;j_3 \lambda_3)
j_{23} \lambda_{23}) j \lambda
\!\mid((j_1 \lambda_1;j_2 \lambda_2)j_{12} \lambda_{12};
j_3 \lambda_3) j \lambda] \nonumber \\
& &=(-1)^{(I_{123}+1)I_{23}+I_{12}\lambda_3+I_{23}\lambda_1+
[j_1+j_2+j_3+j]}
\left\{\begin{array}{ccc}
j_1\lambda_1 & j_2\lambda_2 & j_{12}\lambda_{12} \\
j_3\lambda_3 & j\lambda & j_{23}\lambda_{23}
\end{array}\right\}\,.
\end{eqnarray}
We use the notation $I_{ij...k}=2(j_i+j_j+...+j_k+j_{ij..k})$ ( mod 2 ).
Note that sometimes we use $j$ instead of $j_{123}$ or $j_{1234}$ but
the meaning is always obvious from the context.

Finally, the {\em parity independent super-rotation} $S6-j$ {\em symbols}
[5] are defined by
\begin{eqnarray}
\left\{\begin{array}{ccc}
j_1 & j_2 & j_{12} \\
j_3 & j & j_{23}
\end{array}\right\}^S
= (-1)^{\Psi (\lambda_1,\lambda_2,\lambda_3)}
\left\{\begin{array}{ccc}
j_1\lambda_1 & j_2\lambda_2 & j_3\lambda_3 \\
j_4\lambda_4 & j_5\lambda_5 & j_6\lambda_6
\end{array}\right\}
\end{eqnarray}
where the phase $\Psi$ depends on three independent parities
$\lambda_1,\lambda_2,\lambda_3$ in the following way
\begin{equation}
\Psi(\lambda_1,\lambda_2,\lambda_3) =
(\lambda_1+2j_1)I_{23}+(\lambda_2+2j_2)(I_{123}+I_{12}+I_{23})
+(\lambda_3+2j_3)I_{12}\,,
\end{equation}
The super-rotation $S6-j$ symbols satisfy the
pseudo-orthogonality relations
\begin{equation}
\sum_{j_{12}}(-1)^{I_{12}(I_{123}+1)}
 \left\{ \begin{array}{ccc}
j_1 & j_2 & j_{12}\\
j_3 & j   & j_{23}
\end{array}\right\}^S~
 \left\{ \begin{array}{ccc}
j_1 & j_2 & j_{12}\\
j_3 & j   & j'_{23}
\end{array}\right\}^S
=(-1)^{I_{23}(I_{123}+1)}\delta_{j_{23}\,j^{'}_{23}}~.
\end{equation}
Using the analytical formulae [5] for the $S6-j$ symbols where one of the
super-spins is equal to $\frac{1}{2}$ we obtain a simple
corollary of the pseudo-orthogonality relations (2.16) which almost coincides
with the corresponding equation in the rotation case
\begin{equation}
\sum_{j_{12}}(-1)^{[j_1+j_2+j_{12}+\frac{1}{2}]}
 \left\{ \begin{array}{ccc}
j_1 & j_2 & j_{12}\\
j_2 & j_1 & j'_{12}
\end{array}\right\}^S
=\delta_{j_{12}\,j'_{12}}~.
\end{equation}
\setcounter{equation}{0}

\section{Racah Sum Rule}
\hspace*{10mm}The simplest way of deriving the Racah sum rule and
the Biedenharn-Elliott identity is to use the recoupling theory of,
respectively, three and four superspins, the calculations consisting
in repetitive applications of eqs. (2.11) and (2.12).

In the case of recoupling of three superspins $j_1$, $j_2$ and $j_3$
one can present the transformation (2.12) ( using eq. (2.11) and
summing over the intermediate states containing $j_{13}$ )
as a product of two successive recouplings
 \begin{eqnarray}
& &\mid (j_1 \lambda_1;(j_2 \lambda_2;j_3 \lambda_3)
j_{23} \lambda_{23}) j \lambda;\ell m\rangle \nonumber\\
& &=(-1)^{(\lambda_2+I_{23})(\lambda_3+I_{23})+[j_2+j_3-j_{23}]}
\mid (j_1 \lambda_1;(j_3 \lambda_3;j_2 \lambda_2)
j_{23} \lambda_{23}) j \lambda;\ell m\rangle \nonumber\\
& &=(-1)^{(\lambda_2+I_{23})(\lambda_3+I_{23})+[j_2+j_3-j_{23}]}
\sum_{j_{13}}(-1)^{(\lambda_2+I_{2,13})(\lambda_{13}+I_{2,13})
+[j_2+j_{13}-j]} \nonumber\\
& &\qquad\times[(j_1 \lambda_1;(j_3 \lambda_3;j_2 \lambda_2)
j_{23} \lambda_{23}) j \lambda
\!\mid((j_1 \lambda_1;j_3 \lambda_3)j_{13} \lambda_{13};
j_2 \lambda_2) j \lambda] \nonumber\\
& &\qquad\times
\mid (j_2 \lambda_2;(j_1 \lambda_1;j_3 \lambda_3)j_{13} \lambda_{13})
j \lambda;\ell m\rangle \nonumber\\
& &=(-1)^{(\lambda_2+I_{23})(\lambda_3+I_{23})+[j_2+j_3-j_{23}]}
\sum_{j_{13}}(-1)^{(\lambda_2+I_{2,13})(\lambda_{13}+I_{2,13})
+[j_2+j_{13}-j]} \nonumber\\
& &\qquad\times[(j_1 \lambda_1;(j_3 \lambda_3;j_2 \lambda_2)
j_{23} \lambda_{23}) j \lambda
\!\mid((j_1 \lambda_1;j_3 \lambda_3)j_{13} \lambda_{13};
j_2 \lambda_2) j \lambda] \nonumber\\
& &\qquad\times\sum_{j_{12}}(-1)^{(\lambda_1+I_{12})(\lambda_2+I_{12})
+[j_1+j_2-j_{12}]} \nonumber\\
& &\qquad\times[(j_2 \lambda_2;(j_1 \lambda_1;j_3 \lambda_3)
j_{13} \lambda_{13}) j \lambda
\!\mid((j_2 \lambda_2;j_1 \lambda_1)j_{12} \lambda_{12};
j_3 \lambda_3) j \lambda] \nonumber\\
& &\qquad\times
\mid ((j_1 \lambda_1;j_2 \lambda_2)j_{12} \lambda_{12};j_3 \lambda_3)
j \lambda;\ell m\rangle
 \end{eqnarray}
which gives us the Racah sum rule for the super-rotation recoupling
coefficients in the form
 \begin{eqnarray}
& &[(j_1 \lambda_1;(j_2 \lambda_2;j_3 \lambda_3)
j_{23} \lambda_{23}) j \lambda
\!\mid((j_1 \lambda_1;j_2 \lambda_2)j_{12} \lambda_{12};
j_3 \lambda_3) j \lambda] \nonumber\\
& &=\sum_{j_{13}}(-1)^{\Theta^{R}_{RC}}[(j_1 \lambda_1;
(j_3 \lambda_3;j_2 \lambda_2)
j_{23} \lambda_{23}) j \lambda
\!\mid((j_1 \lambda_1;j_3 \lambda_3)j_{13} \lambda_{13};
j_2 \lambda_2) j \lambda] \nonumber\\
& &\qquad\times[(j_2 \lambda_2;(j_1 \lambda_1;j_3 \lambda_3)
j_{13} \lambda_{13}) j \lambda
\!\mid((j_2 \lambda_2;j_1 \lambda_1)j_{12} \lambda_{12};
j_3 \lambda_3) j \lambda].
 \end{eqnarray}
The phase $\Theta^{R}_{RC}$
 \begin{eqnarray}
& &\Theta^{R}_{RC}=(\lambda_2+I_{23})(\lambda_3+I_{23})+
[j_2+j_3-j_{23}]\nonumber\\
& &\qquad+(\lambda_2+I_{2,13})(\lambda_{13}+I_{2,13})+[j_2+j_{13}-j]+
(\lambda_1+I_{12})(\lambda_2+I_{12})+[j_1+j_2-j_{12}] \nonumber\\
& &=\lambda_1(I_{12}+I_{13}+I_{123})+\lambda_2(I_{12}+I_{23}+I_{123})+
\lambda_3(I_{13}+I_{23}+I_{123})\nonumber\\
& &\qquad+[j_1+j_2-j_{12}]+[j_2+j_3-j_{23}]+[j_2+j_{13}-j]+
I_{12}+I_{23}+I_{123}+I_{123}I_{13}
 \end{eqnarray}
takes the values $0$ or $1$ ( mod $2$ ).
In terms of the parity-dependent $S6-(j,\lambda)$ symbols eq. (3.2) reads
 \begin{equation}
\left\{\begin{array}{ccc}
j_1\lambda_1 & j_2\lambda_2 & j_{12}\lambda_{12} \\
j_3\lambda_3 & j\lambda & j_{23}\lambda_{23}
\end{array}\right\}=
\sum_{j_{13}}(-1)^{\Theta^{R}_{\lambda}}
\left\{\begin{array}{ccc}
j_1\lambda_1 & j_3\lambda_3 & j_{13}\lambda_{13} \\
j_2\lambda_2 & j\lambda & j_{23}\lambda_{23}
\end{array}\right\}
\left\{\begin{array}{ccc}
j_2\lambda_2 & j_1\lambda_1 & j_{12}\lambda_{12} \\
j_3\lambda_3 & j\lambda & j_{13}\lambda_{13}
\end{array}\right\}\,.
\end{equation}
where
\begin{eqnarray}
& &\Theta^{R}_{\lambda}=\Theta_{RC}+(I_{123}+1)I_{13}+
[j_1+j_2+j_3+j] \nonumber\\
& &=\lambda_1(I_{12}+I_{13}+I_{123})+\lambda_2(I_{12}+I_{23}+I_{123})+
\lambda_3(I_{13}+I_{23}+I_{123})+I_{12}+I_{23}+I_{13}\nonumber\\
& &\qquad+I_{123}+[j_1+j_2-j_{12}]+[j_2+j_3-j_{23}]+[j_2+j_{13}-j]+
[j_1+j_2+j_3+j]\,.
\end{eqnarray}

{}From (3.4) using eq.(2.14) we obtain the Racah sum rule for the
super-rotation
$S6-j$ symbols
 \begin{equation}
\left\{\begin{array}{ccc}
j_1 & j_2 & j_{12} \\
j_3 & j   & j_{23}
\end{array}\right\}^S=
\sum_{j_{13}}(-1)^{\Theta^{R}}
\left\{\begin{array}{ccc}
j_1 & j_3 & j_{13} \\
j_2 & j   & j_{23}
\end{array}\right\}^S~
\left\{\begin{array}{ccc}
j_2 & j_1 & j_{12} \\
j_3 & j   & j_{13}
\end{array}\right\}^S\,.
\end{equation}
where
\begin{eqnarray}
& &\Theta^{R}=
[j_1+j_2-j_{12}]+[j_2+j_3-j_{23}]+[j_2+j_{13}-j]+[j_1+j_2+j_3+j] \nonumber\\
& &\qquad+2j_{12}I_{12}+2j_{23}I_{23}+2j_{13}I_{13}+2jI_{123}\,.
\end{eqnarray}

The sign factor in this sum rule is more involved
than the corresponding factor in the ordinary rotation case. However,
they almost coincide in several particular cases. For example, when
the values of the super-spins in the left-hand side of eq. (3.6) are
such that their sums in all the triangles $(j_1,j_2,j_{12})$,
$(j_2,j_3,j_{23})$, $(j_1,j_{23},j)$ and $(j_{12},j_3,j)$ are
integer, Racah sum rule takes the form
 \begin{equation}
\left\{\begin{array}{ccc}
j_1 & j_2 & j_{12} \\
j_3 & j   & j_{23}
\end{array}\right\}^S=
\sum_{j_{13}}(-1)^{[j_{12}+j_{23}+j_{13}]+2j_{13}I_{13}}
\left\{\begin{array}{ccc}
j_1 & j_3 & j_{13} \\
j_2 & j   & j_{23}
\end{array}\right\}^S~
\left\{\begin{array}{ccc}
j_2 & j_1 & j_{12} \\
j_3 & j   & j_{13}
\end{array}\right\}^S\,.
\end{equation}
which means that the terms in the right-hand side, for which the
triangle sums with the participation of $j_{13}$ are also integer,
have the sign factor $(-1)^{j_{12}+j_{23}+j_{13}}$, identical to
the sign factor in the rotation case.

Another particular case of Racah sum rule reads
\begin{equation}
\sum_{j_{12}}(-1)^{2(2(j_1+j_2)(j_{12}+j'_{12})+j_{12})}
 \left\{ \begin{array}{ccc}
j_1 & j_2 & j_{12}\\
j_1 & j_2 & j'_{12}
\end{array}\right\}^S
=1~.
\end{equation}
\setcounter{equation}{0}

\section{Biedenharn-Elliott Identity}
\hspace*{10mm}In order to derive the Biedenharn-Elliott identity we will
consider the recoupling of four super-spins $j_1, j_2, j_3$ and $j_4$.
Using the same technique as in the derivation of eq. (3.2) we can express
the vector $\mid ((j_2 \lambda_2;j_3 \lambda_3)j_{23} \lambda_{23};
(j_1 \lambda_1;j_4 \lambda_4)j_{14} \lambda_{14}) j \lambda;\ell m\rangle$
first, as a result of two successive recouplings of three super-spins, and
second, as a result of three recouplings summing over the intermediate states
containing $j_{124}$. The calculations being completely analogous to the ones
presented in the previous section, we will omit the intermediate (~rather
tedious) calculations and will present only the results: the
Biedenharn-Elliott identity for the super-rotation
recoupling coefficients reads
 \begin{eqnarray}
&&[((j_2 \lambda_2;j_3 \lambda_3)
j_{23} \lambda_{23};(j_1 \lambda_1;j_4 \lambda_4)
j_{14} \lambda_{14}) j \lambda
\!\mid(((j_2 \lambda_2;j_3 \lambda_3)
j_{23} \lambda_{23};j_1 \lambda_1)j_{123} \lambda_{123};
j_4 \lambda_4) j \lambda] \nonumber\\
&&\times[((j_1 \lambda_1;(j_2 \lambda_2;j_3 \lambda_3)
j_{23} \lambda_{23})j_{123} \lambda_{123};
j_4 \lambda_4) j \lambda
\!\mid(((j_1 \lambda_1;j_2 \lambda_2)
j_{12} \lambda_{12};j_3 \lambda_3)j_{123} \lambda_{123};
j_4 \lambda_4) j \lambda]\nonumber\\
&&=\sum_{j_{124}}(-1)^{\Theta^{BE}_{RC}}\nonumber\\
&&\times[((j_1 \lambda_1;
j_4 \lambda_4)j_{14} \lambda_{14};(j_2 \lambda_2;j_3 \lambda_3)
j_{23} \lambda_{23}) j \lambda
\!\mid(((j_1 \lambda_1;j_4 \lambda_4)j_{14} \lambda_{14};
j_2 \lambda_2)j_{124} \lambda_{124};j_3 \lambda_3) j \lambda]\nonumber\\
&&\times[((j_2 \lambda_2;(j_1 \lambda_1;j_4 \lambda_4)
j_{14} \lambda_{14})j_{124} \lambda_{124};j_3 \lambda_3) j \lambda
\!\mid(((j_2 \lambda_2;j_1 \lambda_1)j_{12} \lambda_{12};
j_4 \lambda_4)j_{124} \lambda_{124};j_3 \lambda_3)j \lambda]\nonumber\\
&&\times[(j_3 \lambda_3;((j_1 \lambda_1;j_2 \lambda_2)j_{12} \lambda_{12};
j_4 \lambda_4)j_{124} \lambda_{124})j \lambda
\!\mid((j_3 \lambda_3;(j_1 \lambda_1;j_2 \lambda_2)
j_{12} \lambda_{12})j_{123} \lambda_{123};
j_4 \lambda_4) j \lambda].\nonumber\\
&&\mbox{}
 \end{eqnarray}
It is interesting to note that the phase $\Theta^{BE}_{RC}$ is
independent of the parities
 \begin{eqnarray}
&&\Theta^{BE}_{RC}=I_{23}(I_{1234}+I_{123}+I_{14})+
I_{124}(I_{1234}+I_{14}+1)+I_{12}(I_{123}+1)+I_{1234}I_{14}\nonumber\\
&&+[j_1+j_2-j_{12}]+[j_{12}+j_3-j_{123}]+[j_1+j_{23}-j_{123}]\nonumber\\
&&+[j_{14}+j_2-j_{124}]+[j_{14}+j_{23}-j]+[j_{124}+j_3-j]\,.
 \end{eqnarray}
As a result, the Biedenharn-Elliott identities for the $\lambda$-dependent
$S6-(j,\lambda)$ and $\lambda$-indepen\-dent super-rotation $S6-j$ symbols
take the same form
 \begin{eqnarray}
&&\left\{\begin{array}{ccc}
j_1 & j_2 & j_{12} \\
j_3 & j_{123} & j_{23}
\end{array}\right\}^S
\left\{\begin{array}{ccc}
j_{23} & j_1 & j_{123} \\
j_4 & j & j_{14}
\end{array}\right\}^S \nonumber\\
&&=\sum_{j_{124}}(-1)^{\Theta^{BE}}
\left\{\begin{array}{ccc}
j_2 & j_1 & j_{12} \\
j_4 & j_{124} & j_{14}
\end{array}\right\}^S~
\left\{\begin{array}{ccc}
j_3 & j_{12} & j_{123} \\
j_4 & j   & j_{124}
\end{array}\right\}^S~
\left\{\begin{array}{ccc}
j_{14} & j_2 & j_{124}\\
j_3 & j   & j_{23}
\end{array}\right\}^S\,.
\end{eqnarray}
where
\begin{eqnarray}
&&\Theta^{BE}=I_{12}(I_{1234}+I_{124}+I_{123}+1)+I_{14}I_{23}\nonumber\\
&&\qquad+[j_1+j_2-j_{12}]+[j_{12}+j_3-j_{123}]+[j_1+j_{23}-j_{123}]\nonumber\\
&&\qquad+[j_{14}+j_2-j_{124}]+[j_{14}+j_{23}-j]+[j_{124}+j_3-j]\nonumber\\
&&\qquad+[j_1+j_2+j_3+j_{123}]+[j_1+j_4+j_{23}+j]+
[j_1+j_2+j_4+j_{124}]\nonumber\\
&&\qquad+[j_3+j_4+j_{12}+j]+[j_2+j_3+j_{14}+j]\,.
\end{eqnarray}

Here, as well as in the case of the Racah sum rule, we see that the structure
of the Biedenharn-Elliott identity for the super-rotation $S6-j$ symbols
is exactly the same as the structure of the Biedenharn-Elliott identity [8,9]
for the rotation $6-j$ symbols. Only the sign factor $\Theta^{BE}$
is somewhat more involved in the supersymmetric case. This factor
almost coincides with the corresponding one in the rotation case when the
values of the super-spins in the left-hand side of eq. (4.3) are
such that their sums in all the triangles $(j_1,j_2,j_{12})$,
$(j_2,j_3,j_{23})$, $(j_1,j_{23},j_{123})$, $(j_{12},j_3,j_{123})$,
$(j_1,j_4,j_{14})$, $(j_{123},j_4,j)$ and $(j_{14},j_{23},j)$ are
integer
 \begin{eqnarray}
&&\left\{\begin{array}{ccc}
j_1 & j_2 & j_{12} \\
j_3 & j_{123} & j_{23}
\end{array}\right\}^S
\left\{\begin{array}{ccc}
j_{23} & j_1 & j_{123} \\
j_4 & j & j_{14}
\end{array}\right\}^S \nonumber\\
&&=\sum_{j_{124}}(-1)^{[j_1+j_2+j_3+j_4+j_{12}+j_{23}+j_{14}+
j_{123}+j_{124}+j+\frac{1}{2}]} \nonumber\\
&&\qquad\times\left\{\begin{array}{ccc}
j_2 & j_1 & j_{12} \\
j_4 & j_{124} & j_{14}
\end{array}\right\}^S~
\left\{\begin{array}{ccc}
j_3 & j_{12} & j_{123} \\
j_4 & j   & j_{124}
\end{array}\right\}^S~
\left\{\begin{array}{ccc}
j_{14} & j_2 & j_{124}\\
j_3 & j   & j_{23}
\end{array}\right\}^S\,.
\end{eqnarray}

\section{Conclusion}
\hspace*{10mm}In this paper we continued the development of the
Racah-Wigner calculus for the super-rotation $osp(1|2)$
superalgebra. In particular, we have derived the Racah sum rule and the
Biedenharn-Elliott identity for the super-rotation $S6-j$ symbols.
It turned out that the structure of these relations is the same as
the corresponding structure in the rotation case. Only the sign factors are
somewhat more involved.

In a forthcoming publication we will use the Biedenharn-Elliott
identity in order to obtain recursion relations for the
super-rotation $S6-j$ symbols. These relations will permit us to develop
numerical algorithms for the evaluation of the $S6-j$ symbols
for large values of the superspins, our final goal being to
obtain a Regge-Ponzano [10] type formula for the $S6-j$ symbols.

\section*{Acknowledgements}
One of the authors (S.T.) was supported by grant from The Commission
of the European Communities under contract $n^o$ ERB3510PL920706 706
( ERB-CIPA-CT-92-2137 proposal NR.: 706 ). He
acknowledges the warm hospitality of the Laboratoire de Physique
Th\'{e}orique de l'Universit\'{e} de Bordeaux I.

\newpage
\section*{References}

\noindent
[1] A. Pais and V. Rittenberg, J. Math. Phys. {\bf 16}, 2062 (1975).
\newline
\noindent
[2] M. Scheunert, W. Nahm, and V. Rittenberg,	J. Math. Phys. {\bf 18}, 146
and 155 (1977).
\newline
\noindent
[3] F. A. Berezin and V. N. Tolstoy, Commun. Math. Phys. {\bf 78}, 409 (1981).
\newline
\noindent
[4] P. Minnaert and M. Mozrzymas, J. Math. Phys. {\bf 33}, 1582
and 1594 (1992).
\newline
\noindent
[5] M. Daumens, P. Minnaert, M. Mozrzymas and S. Toshev, J. Math. Phys.
{\bf 34}, 2475 (1993).
\newline
\noindent
[6] M. Daumens, P. Minnaert, M. Mozrzymas and S. Toshev, Europhys. Lett.
{\bf 20}, 671 (1992).
\newline
\noindent
[7] L. C. Biedenharn and J. S. Louck, {\em Angular Momentum in
Quantum Physics; Encyclopedia of Mathematics and Its Applications},
Addison-Wesley, London, (1981), Vol. 8, p. 106.
\newline
\noindent
[8] L. C. Biedenharn, J. Math. and Phys. {\bf 31}, 287 (1953).
\newline
\noindent
[9] J. P. Elliott, Proc. Roy. Soc. {\bf A218}, 345 (1953).
\newline
\noindent
[10] G. Ponzano and T. Regge, in: {\em Spectroscopic and Group Theoretical
Methods in Physics (Racah Memorial Volume)}, eds. F.Bloch {\em et al.},
North-Holland, Amsterdam, (1968), p. 1.
\end{document}